\documentclass[aps,prl,reprint,twocolumn,superscriptaddress]{revtex4-2}

\usepackage{dcolumn}
\usepackage{bm}
\usepackage[utf8]{inputenc}
\usepackage[english]{babel}
\usepackage{amsfonts}
\usepackage{amssymb}
\usepackage{graphicx}
\usepackage{setspace}
\usepackage{verbatim}
\usepackage{xcolor}
\usepackage{hyperref}
\usepackage{mathrsfs}
\usepackage{cancel}
\usepackage{amsmath}
\usepackage{braket}
\usepackage{color}
\usepackage{soul}
\usepackage{slashed}
\usepackage[normalem]{ulem}

\newcommand{\be}{\begin{equation}}
\newcommand{\ee}{\end{equation}}
\newcommand{\bea}{\begin{eqnarray}}
\newcommand{\eea}{\end{eqnarray}}

\colorlet{RED}{red}
\colorlet{ORANGE}{orange}
\colorlet{BLUE}{blue}
\colorlet{PURPLE}{purple}

\newcommand{\TT}{T^2}
\newcommand{\TTa}{T^2_\alpha}%

\begin{document}

\title{Exact stress-tensor fluctuations from conformal symmetry 
\\ and black hole radiation
}

\author{Jos\'e Navarro-Salas}
\email{jnavarro@ific.uv.es}
\affiliation{Instituto de F\'isica Corpuscular (IFIC), CSIC-Universitat de Val\`encia and Departament de F\'isica Te\`orica, Facultad de F\'isica,  Burjassot-46100, Valencia, Spain}

\author{Roberto Balbinot}
\email{roberto.balbinot@unibo.it}
\affiliation{Dipartimento di Fisica e Astronomia ``Augusto Righi'', Universit\`a di Bologna, Via Irnerio 46, 40126 Bologna, Italy}
\affiliation{INFN Sezione di Bologna, Viale Berti Pichat 6/2, 40127 Bologna, Italy}

\author{Alessandro Fabbri}
\email{afabbri@ific.uv.es}
\affiliation{Instituto de F\'isica Corpuscular (IFIC), CSIC-Universitat de Val\`encia and Departament de F\'isica Te\`orica, Facultad de F\'isica,  Burjassot-46100, Valencia, Spain}

\begin{abstract}
Semiclassical gravity is sourced by the expectation value of the stress tensor, and it is reliable only if the fluctuations around this value are small. In two dimensions, conformal symmetry captures the essential physics of Hawking radiation in its s-wave sector and determines the mean value exactly, for any background and any quantum state. We show that it also determines the local fluctuations of the null components $T_{\pm\pm}$. The square of the stress tensor is defined by the operator product expansion, up to a multiple $\alpha$ of $\partial_\pm^2T_{\pm\pm}$ that labels point-splitting prescriptions. Its transformation law under the conformal map to locally inertial coordinates gives the variance in closed form. The relative fluctuation scales as $1/c$, with $c$ the central charge. For the Schwarzschild black hole it is $22/5c$ for the Hawking flux at infinity, that of flat-space thermal radiation. At the horizon it is $-44\alpha/3c$ for the outgoing flux and $(238+720\alpha)/5c$ for the ingoing flux that drives the backreaction. Both reduce to $22/5c$ for $\alpha=-3/10$, which corresponds to the M\"obius-covariant prescription. For large $c$ these results support the semiclassical approximation.
\end{abstract}

\date{\today}

\maketitle

\textbf{Introduction}.---Since Hawking's discovery that black holes emit thermal radiation \cite{Hawking75}, the semiclassical approximation has been the standard framework for analyzing the quantum behavior of black holes and their evaporation \cite{BirrellDavies, Wald94, ParkerToms}. In this framework, quantum matter propagates on a classical background, which responds to the expectation value $\langle T_{\mu\nu}\rangle$ of the renormalized stress-energy tensor. This framework assumes that the quantum fluctuations of the stress tensor around its mean value are small. If they were large, the geometry that backreacts to $\langle T_{\mu\nu}\rangle$ might not describe the actual physics properly. A more refined approach, which includes the fluctuations of the source, might then be needed \cite{HuVerdaguer}.

In this context, one has to quantify the fluctuations around $\langle T_{\mu\nu}\rangle$, and this requires, essentially, the square of the components of the stress tensor. More precisely, the local fluctuations of a component $T_{\mu\nu}$ are measured by its variance $\langle T_{\mu\nu}^2\rangle-\langle T_{\mu\nu}\rangle^2$. This requires a renormalization prescription that goes beyond the one used for the mean value $\langle T_{\mu\nu}\rangle$. The variance should be regarded as the coincidence limit of the connected two-point function of the stress tensor, which is even more divergent in the ultraviolet than the mean value. In flat space the leading divergence is state independent, while the subleading ones depend on the state, and both are removed by normal ordering \cite{KuoFord93, WuFord99}. In a curved background, however, additional geometry-dependent short-distance singularities appear. It is then difficult to construct a well-defined prescription for the local fluctuations \cite{HuVerdaguer, PerezSudarsky26, Paul-Carmen-Emil}.

A natural framework to address this problem is provided by two-dimensional conformal symmetry, which occupies a privileged place in physics \cite{BPZ84, DiFrancesco}.  Its infinite-dimensional Virasoro algebra underlies, to mention only two major examples, the classification of critical phenomena in statistical mechanics \cite{FQS84,Cardy10} and the worldsheet formulation of string theory \cite{Polchinski98}. Remarkably, the same symmetry also captures the essential physics of black hole radiation. When the angular degrees of freedom are frozen, the s-wave sector of a matter field propagating on a black hole is well described by a conformal field theory of central charge $c$ on an effective two-dimensional spacetime, whose metric can always be written in conformal gauge, $ds^2=-e^{2\rho}dx^+dx^-$ \cite{Strominger95,RobinsonWilczek05,FabbriNavarroSalas}. We use units in which $\hbar$, the speed of light, Newton's constant $G$ and Boltzmann's constant $k_B$ are equal to one. Conformal symmetry then determines the stress tensor up to state-dependent terms, and its null components take the exact local form
\begin{equation}\label{eq:Tmean}
\langle T_{\pm\pm}\rangle=-\frac{c}{12\pi}\left[(\partial_\pm\rho)^2-\partial_\pm^2\rho\right]+\langle{:}T_{\pm\pm}{:}\rangle ,
\end{equation}
where $c$ is the central charge of the conformal field theory, which effectively counts the number of matter fields, and only the normal-ordered part $\langle{:}T_{\pm\pm}{:}\rangle$ depends on the quantum state~\cite{noteDFU}. For the Schwarzschild geometry, $e^{2\rho}=1-2M/r$ in terms of the double null Eddington--Finkelstein coordinates $x^-=u$ and $x^+=v$, and in the state describing gravitational collapse Eq.~(\ref{eq:Tmean}) yields, far from the black hole, the outgoing flux
\begin{equation}\label{eq:Hawkingflux}
\langle T_{--}\rangle=\frac{c\kappa^2}{48\pi} ,
\end{equation}
with $\kappa=1/4M$ the surface gravity and $T_H=\kappa/2\pi$ the Hawking temperature. 
This is just the energy flux of the thermal radiation at temperature $T_H$ in two dimensions. It  matches  the s-wave Hawking flux  when the backscattering off the potential barrier is neglected.

Equation~(\ref{eq:Tmean}) gives the mean value at each point in terms of the geometry and of the state. No similar formula is available for the fluctuations in a general two-dimensional conformal field theory \cite{notePS}. In this paper we obtain such a formula. The square of the stress tensor is represented by the normal-ordered product $T_{\pm\pm}^2$ of conformal field theory or, more generally, by $T^2_{\pm\pm}+\alpha\,\partial_{\pm}^2T_{\pm\pm}$, where $\alpha$ labels different point-splitting prescriptions. All terms quadratic in $c$ cancel, so that, for $N=c$ fields in identical states, the relative fluctuations are suppressed as $1/c$, which provides a quantitative justification of the semiclassical large-$N$ limit \cite{Tomboulis77,Strominger95}. Applied to the Schwarzschild black hole, our formula gives the relative fluctuation $22/5c$ for the Hawking flux at infinity, Eq.~(\ref{eq:Hawkingflux}), which is that of thermal radiation in flat spacetime, for any $\alpha$. At the horizon the fluctuations of the outgoing flux, in a regular frame, and of the ingoing flux, which drives the backreaction, depend on $\alpha$.

\textbf{Relevant operator product expansions in conformal field theory}.---We now show how Eq.~(\ref{eq:Tmean}) can be generalized to the second moment of $T_{\pm\pm}$, and hence to the variance. To this end we use standard techniques of conformal field theory \cite{DiFrancesco} and, as usual, work with complex coordinates $(z,\bar z)$ on the Euclidean plane. We shall return to Lorentzian signature at the end.

In this language the holomorphic component of the stress tensor, normalized as usual in conformal field theory, $T(z)=-2\pi T_{zz}$, obeys the operator product expansion
\begin{equation}\label{eq:TTOPE}
T(z)T(w)\sim\frac{c/2}{(z-w)^4}+\frac{2T(w)}{(z-w)^2}+\frac{\partial T(w)}{z-w},
\end{equation}
and under a finite holomorphic map $\omega=f(z)$ it transforms as
\begin{equation}\label{eq:Tfinite}
\widetilde T(\omega)=p^{-2}\left[T(z)-\frac{c}{12}S\right],
\end{equation}
where $p=f'$, $\sigma=f''/f'$, and $S=\sigma'-\sigma^2/2$ is the Schwarzian derivative of $f$. The antiholomorphic component $\bar T(\bar z)$ obeys the same relations with $z\to\bar z$. The inhomogeneous term in Eq.~(\ref{eq:Tfinite}) is the origin of the geometric part of Eq.~(\ref{eq:Tmean}).

The natural local operator measuring the fluctuations is the composite $T^2\equiv(TT)$, the normal-ordered product of conformal field theory \cite{DiFrancesco}. It is the regular part of the operator product $T(z)T(w)$ as $z\to w$, that is, the limit of $T(z)T(w)$ once the singular terms of Eq.~(\ref{eq:TTOPE}), with their coefficients evaluated at $w$, have been subtracted. With this convention $\TT$ is uniquely defined, in any state and for any $c$. The stress tensor is itself a normal-ordered product of this kind, for instance $T=-\frac{1}{2}(\partial\varphi\,\partial\varphi)$ for a free boson, and we do not write the normal-ordering symbol explicitly in this section. Both $T$ and $\TT$ are normal ordered with respect to the coordinate $z$ in which the operator product expansion is written, and their transformation laws relate the operators normal ordered with respect to $z$ and with respect to $\omega$. The generalized Wick theorem gives \cite{DiFrancesco}
\begin{align}\label{eq:TTTOPE}
T(z)\TT(w)\sim{}&\frac{3c}{(z-w)^6}+\frac{(c+8)\,T(w)}{(z-w)^4}+\frac{3\,\partial T(w)}{(z-w)^3}\notag\\
&+\frac{4\,\TT(w)}{(z-w)^2}+\frac{\partial\TT(w)}{z-w}.
\end{align}

The subtraction that defines $\TT$ is a point-splitting prescription, and it is not the only possible one. If the coefficients of the singular terms are evaluated at an intermediate point $w+\lambda(z-w)$, with $0\leq\lambda\leq1$, the finite remainder is $\TT-\lambda(1-\lambda)\,\partial^2T$ (Appendix~\ref{app:A}). It is $\TT$ for $\lambda=0$ and $\lambda=1$. Each prescription defines a different operator, and none of them is ambiguous. Since $\partial^2T$ is the only other operator of weight four that can be built from the stress tensor alone, we shall consider the family
\begin{equation}\label{eq:family}
\TTa\equiv\TT+\alpha\,\partial^2T ,\qquad a\equiv3+10\alpha ,
\end{equation}
where $\alpha$ is a real number that does not depend on $c$.

The transformation law of an operator is determined by the singular terms of its operator product with $T(z)$ \cite{DiFrancesco}, and these depend on $\alpha$. From Eqs.~(\ref{eq:TTOPE}) and (\ref{eq:TTTOPE}), the poles of order six, four and three of $T(z)\TTa(w)$ have the coefficients $c\,a$, $(c+8+12\alpha)\,T$ and $a\,\partial T$, while those of order two and one, $4\,\TTa$ and $\partial\TTa$, are common to all the operators of weight four. As for $T$, the pole of order six gives rise to inhomogeneous c-number terms, and the pole of order three makes $\TTa$ mix with $\partial T$ and $T$. The finite transformation law, derived in Appendix~\ref{app:A}, is
\begin{align}\label{eq:TTfinite}
\widetilde{\TTa}(\omega)={}&p^{-4}\bigg[\TTa-\frac{a}{2}\,\sigma\,\partial T+\frac{3a\sigma^2-(c+8+12\alpha)S}{6}\,T\notag\\
&+\frac{c}{720}\big[(5c+58+120\alpha)S^2\notag\\
&\qquad+30a\,\sigma(S'-\sigma S)-6a\,S''\big]\bigg],
\end{align}
with the fields on the right-hand side evaluated at $z$. Its infinitesimal form reproduces the variation implied by the operator product expansion, and it is consistent under composition. For $\alpha=0$ it is the law of the normal-ordered product. For $a=0$, that is, $\alpha=-3/10$, the poles of order six and three are absent, and Eq.~(\ref{eq:TTfinite}) involves only the Schwarzian derivative. The operator is then the quasi-primary field $\Lambda\equiv\TT-\frac{3}{10}\partial^2T$ \cite{Zamolodchikov85}, in terms of which $\TTa=\Lambda+\frac{a}{10}\partial^2T$. Only for this value does $\TTa$ transform as a tensor under the M\"obius maps, as $T$ does (Appendix~\ref{app:A}).

\textbf{Local expression of $\langle T_{\pm\pm}^2\rangle$ in curved space}.---We now return to the Lorentzian metric $ds^2=-e^{2\rho}dx^+dx^-$. The holomorphic and antiholomorphic sectors become the two chiral sectors, which depend on the null coordinates $x^+$ and $x^-$, and the formulas of the previous section apply to each of them with $z\to x^\pm$ and $\partial\to\partial_\pm$. In the physical normalization of Refs.~\cite{FabbriNavarroSalas,BirrellDavies}, ${:}T_{\pm\pm}{:}=-T/2\pi$, ${:}T_{\pm\pm}^2{:}_\alpha=\TTa/4\pi^2$ and ${:}\Lambda_{\pm\pm}{:}=\Lambda/4\pi^2$. From now on we write the normal ordering explicitly.

Normal ordering depends on the choice of coordinates. We denote by $\langle{:}T_{\pm\pm}(x^\pm){:}\rangle$ and $\langle{:}T_{\pm\pm}^2(x^\pm){:}_\alpha\rangle$ the expectation values normal ordered with respect to the positive-frequency modes with respect to $x^\pm$. They carry all the information on the quantum state and vanish in the vacuum defined by these modes. Physical expectation values are defined by the locally inertial prescription of Refs.~\cite{FabbriNavarroSalas, FFN03}. At each point $X$ one introduces conformal coordinates $\xi_X^\pm(x^\pm)$ that are locally inertial at $X$, normal orders with respect to them, and transforms back to the coordinates $x^\pm$,
\begin{align}\label{eq:prescription}
\langle T_{\pm\pm}(X)\rangle&=p_\pm^2\,\langle{:}T_{\pm\pm}(\xi_X^\pm){:}\rangle\big|_X ,\notag\\
\langle T_{\pm\pm}^2(X)\rangle_\alpha&=p_\pm^4\,\langle{:}T_{\pm\pm}^2(\xi_X^\pm){:}_\alpha\rangle\big|_X ,
\end{align}
with $p_\pm=d\xi_X^\pm/dx^\pm$. The coordinates $\xi_X^\pm$ must remove at $X$ the Christoffel symbols $\Gamma^\pm_{\pm\pm}=2\partial_\pm\rho$ and their chiral derivatives. They are defined by
\begin{equation}\label{eq:xiX}
\frac{d\xi_X^\pm}{dx^\pm}=b_\pm\,e^{2[\rho(x^\pm,x^\mp(X))-\rho(X)]} ,
\end{equation}
with $b_+b_-=e^{2\rho(X)}$, so that along each null line through $X$ the coordinate $\xi_X^\pm$ is an affine parameter. The mixed derivative $\partial_+\partial_-\rho$, proportional to the scalar curvature, cannot be removed by a conformal map. At $X$ the Schwarzian derivative of the map $x^\pm\to\xi_X^\pm$ is $S_\pm=2\rho_2-2\rho_1^2$, with $\rho_n=\partial_\pm^n\rho(X)$.

For the stress tensor, Eq.~(\ref{eq:Tfinite}) with $z=x^\pm$ and $\omega=\xi_X^\pm$ gives $\langle T_{\pm\pm}\rangle=\langle{:}T_{\pm\pm}{:}\rangle+cS_\pm/24\pi$, which is Eq.~(\ref{eq:Tmean}). Proceeding in the same way with Eq.~(\ref{eq:TTfinite}), and subtracting $\langle T_{\pm\pm}\rangle^2$, we arrive at (Appendix~\ref{app:B})
\begin{align}\label{eq:variance}
&\langle T_{\pm\pm}^2\rangle_\alpha-\langle T_{\pm\pm}\rangle^2=\langle{:}\Lambda_{\pm\pm}{:}\rangle-\langle{:}T_{\pm\pm}{:}\rangle^2\notag\\
&\quad+\frac{11}{15\pi}\big(\rho_2-\rho_1^2\big)\langle{:}T_{\pm\pm}{:}\rangle+\frac{11c}{360\pi^2}\big(\rho_2-\rho_1^2\big)^2\notag\\
&\quad-\frac{a}{20\pi}\,\nabla_\pm\nabla_\pm\langle T_{\pm\pm}\rangle ,
\end{align}
where ${:}\Lambda_{\pm\pm}{:}={:}T_{\pm\pm}^2{:}_\alpha+\frac{a}{20\pi}\partial_\pm^2{:}T_{\pm\pm}{:}$, $\langle T_{\pm\pm}\rangle$ is given by Eq.~(\ref{eq:Tmean}), and $\nabla_\pm\nabla_\pm\langle T_{\pm\pm}\rangle=[\partial_\pm^2-10\rho_1\partial_\pm+24\rho_1^2-4\rho_2]\langle T_{\pm\pm}\rangle$ is its second covariant derivative. Equation~(\ref{eq:variance}), the analogue of Eq.~(\ref{eq:Tmean}) for the fluctuations, is one of the main results of this paper. It is local, and it holds for any quantum state and any $\alpha$. The terms quadratic in $c$ have cancelled, and the parameter $\alpha$ enters only through the last term, which is linear in the mean value. For the normal-ordered product, $\alpha=0$, the state-independent part is $\frac{c}{720\pi^2}(238\rho_1^4-476\rho_1^2\rho_2+76\rho_2^2+108\rho_1\rho_3-9\rho_4)$.

In flat space, $\rho=0$, Eq.~(\ref{eq:variance}) reduces to the normal-ordered connected part $\langle{:}T_{\pm\pm}^2{:}_\alpha\rangle-\langle{:}T_{\pm\pm}{:}\rangle^2$. Moreover, if the state is a product of identical states for $N=c$ free fields, both $\langle{:}T_{\pm\pm}{:}\rangle$ and $\langle{:}T_{\pm\pm}^2{:}_\alpha\rangle-\langle{:}T_{\pm\pm}{:}\rangle^2$ are linear in $c$, and so is the whole right-hand side of Eq.~(\ref{eq:variance}). Since $\langle T_{\pm\pm}\rangle^2$ grows as $c^2$, the relative fluctuation scales as
\begin{equation}\label{eq:ratio}
\frac{\langle T_{\pm\pm}^2\rangle_\alpha-\langle T_{\pm\pm}\rangle^2}{\langle T_{\pm\pm}\rangle^2}=\frac{1}{c}\,\frac{\langle T_{\pm\pm}^2\rangle_{\alpha(1)}-\langle T_{\pm\pm}\rangle_{(1)}^2}{\langle T_{\pm\pm}\rangle_{(1)}^2},
\end{equation}
where the subscript $(1)$ refers to a single field. The dependence on $c$ is exactly $1/c$ for every $\alpha$. The fluctuations are therefore suppressed only for large $c$, which justifies the semiclassical approximation in that limit. This agrees with the general result that, in the geometric quantization of the Virasoro group, $c^{-1}$ plays the role of Planck's constant, so that large $c$ is the classical limit~\cite{Witten88}. This clarifies the validity of the semiclassical approximation, which was questioned in Ref.~\cite{PerezSudarsky26}.

The result is particularly simple for the vacuum state associated with an arbitrary choice of null coordinates, such as the Boulware, Unruh and Hartle--Hawking states. Equation~(\ref{eq:variance}) then becomes (Appendix~\ref{app:B})
\begin{equation}\label{eq:covvac}
\langle T_{\pm\pm}^2\rangle_\alpha-\langle T_{\pm\pm}\rangle^2=\frac{22}{5c}\,\langle T_{\pm\pm}\rangle^2-\frac{a}{20\pi}\,\nabla_\pm\nabla_\pm\langle T_{\pm\pm}\rangle ,
\end{equation}
at every point and in any background. For $a=0$ the relative fluctuation is the constant $22/5c$, that of thermal radiation in flat space.

For $N=c$ free scalar fields one could instead take the square of the total stress tensor with all self-contractions removed, which extends to $c$ fields the definition adopted in Ref.~\cite{PerezSudarsky26} for a single field. This operator is not a member of the family (\ref{eq:family}), and it does not depend on the point-splitting prescription. For vacuum states it gives the relative fluctuation $2/c$ at every point (Appendix~\ref{app:W}). The operators $\TTa$ are built from the stress tensor alone and are defined for any value of $c$. 

\textbf{Schwarzschild black hole: fluctuations of the Hawking flux}.---%
Let us now return to the Schwarzschild black hole, with the double null coordinates $(u,v)$ introduced above, where $(v-u)/2=r^*=r+2M\ln|r/2M-1|$.
In the Unruh state $|U\rangle$ \cite{unruh76}, which describes the late-time Hawking radiation following from gravitational collapse, the outgoing modes are positive-frequency modes with respect to the Kruskal coordinate $U=-\frac{1}{\kappa}e^{-\kappa u}$, while the ingoing modes are positive-frequency modes with respect to $v$, as in the Boulware state. This means $\langle U|{:}T_{UU}{:}|U\rangle=0$ and, from Eq.~(\ref{eq:Tfinite}), with $\omega=U$ and $z=u$, $\langle U|{:}T_{uu}{:}|U\rangle=\frac{c\kappa^2}{48\pi}$. Equation~(\ref{eq:covvac}) then gives the Unruh state variance
\begin{align}\label{eq:varianceU}
&\langle U|T_{uu}^2|U\rangle_\alpha-\langle U|T_{uu}|U\rangle^2=\frac{c\,(r-2M)^4\,N_\alpha(r)}{1474560\,\pi^2M^4r^8},\notag\\
&N_\alpha\equiv(r-2M)\big(11r^3+110Mr^2+660M^2r+1800M^3\big)\notag\\
&\qquad\quad-1920\,\alpha\,M^3(r+9M),
\end{align}
while the mean value is $\langle U|T_{uu}|U\rangle=c\,(r-2M)^2(r^2+4Mr+12M^2)/768\pi M^2r^4$ \cite{FabbriNavarroSalas}.

Far from the black hole the variance tends to the constant value $\frac{11c\kappa^4}{5760\pi^2}$, the same result as for thermal radiation at the temperature $T_H$ in flat space, obtained directly from the thermal two-point function (Appendix~\ref{app:T}). [The Wick-ordered square of Ref.~\cite{PerezSudarsky26} does not reproduce this value. It gives $\frac{5c\kappa^4}{5760\pi^2}$.]  Together with Eq.~(\ref{eq:Hawkingflux}), it gives the relative fluctuation $22/5c$ of the Hawking flux at infinity, for any $\alpha$. It is the local counterpart of the thermodynamic fluctuation formula, with the same dependence on $c$ (Appendix~\ref{app:T}).

Near the horizon the result depends on $\alpha$. The variance (\ref{eq:varianceU}) vanishes there as $(r-2M)^4$, which is the behavior required for a finite value in a frame regular at the future horizon. In the Kruskal coordinate $U$, where $\frac{du}{dU}\simeq e^{\kappa v-1}\frac{2M}{r-2M}$, we obtain on the horizon
\begin{equation}\label{eq:0variance}
\langle U|T_{UU}^2|U\rangle_\alpha-\langle U|T_{UU}|U\rangle^2\big|_{r=2M}=-\frac{11\,\alpha\,c\kappa^4}{48\pi^2}\,e^{4(\kappa v-1)} ,
\end{equation}
while $\langle U|T_{UU}|U\rangle|_{r=2M}=\frac{c\kappa^2}{8\pi}e^{2(\kappa v-1)}$. The relative fluctuation of the outgoing flux at the horizon is therefore $-44\alpha/3c$. For the normal-ordered product, $\alpha=0$, the variance vanishes on the horizon and changes sign there. Since the Unruh state and the coordinate $U$ are regular across the future horizon, Eq.~(\ref{eq:varianceU}) also holds for $r<2M$, where the variance is then negative. For $\alpha<0$ it is positive at the horizon and everywhere outside. For $\alpha=-3/10$, $N_\alpha=11(r^2+4Mr+12M^2)^2$ and the relative fluctuation is $22/5c$ at every point.

\textbf{Fluctuations of the ingoing flux near the horizon}.---
We can also apply our general formula to the ingoing quantum flux near the horizon, which drives, via the backreaction equations, the evolution of the horizon \cite{Bardeen, York}. To this end one should evaluate $\langle U|T_{vv}^{EF}|U\rangle$ in ingoing Eddington--Finkelstein coordinates $(v,r)$, where the metric takes the form $ds^2= -(1-2M/r)dv^2 +2dvdr$, and $T_{vv}^{EF}=T_{uu}+T_{vv}+2T_{uv}$. For a conformal field theory the trace is proportional to the identity operator, $T_{uv}=-\frac{c}{12\pi}\partial_u\partial_v\rho$, and it does not fluctuate. The two chiral sectors are uncorrelated in the Unruh state, where, as usual, contact terms at coincident points are discarded. The variance of $T_{vv}^{EF}$ is then the sum of the two chiral variances (Appendix~\ref{app:B}),
\begin{align}\label{eq:varEF}
\langle (T_{vv}^{EF})^2\rangle_\alpha-\langle T_{vv}^{EF}\rangle^2={}&\big(\langle T_{uu}^2\rangle_\alpha-\langle T_{uu}\rangle^2\big)\notag\\
&+\big(\langle T_{vv}^2\rangle_\alpha-\langle T_{vv}\rangle^2\big).
\end{align}
On the horizon the outgoing variance, Eq.~(\ref{eq:varianceU}), vanishes for any $\alpha$, as do $\langle T_{uu}\rangle$ and $T_{uv}$, and only the ingoing sector contributes. For it $\langle U|{:}T_{vv}{:}|U\rangle=0$, and the variance follows from Eq.~(\ref{eq:covvac}). The mean value is the well-known negative ingoing flux $\langle T_{vv}^{EF}\rangle|_{r=2M}=-c\kappa^2/48\pi$, and the relative fluctuation is
\begin{equation}\label{eq:RinfluxH}
\frac{\langle (T_{vv}^{EF})^2\rangle_\alpha-\langle T_{vv}^{EF}\rangle^2}{\langle T_{vv}^{EF}\rangle^2}\bigg|_{r=2M}=\frac{238+720\,\alpha}{5c}.
\end{equation}
For the normal-ordered product this is $238/5c\simeq47.6/c$, about eleven times the value $22/5c$ of the Hawking flux at infinity. For $\alpha=-3/10$ it is $22/5c$, and the variance, $11c\kappa^4/5760\pi^2$, is exactly that of the Hawking flux at infinity.

\textbf{Boulware vacuum and the dependence on $\alpha$}.---In the Boulware state, the vacuum associated with the coordinates $(u,v)$, which describes the vacuum polarization outside a static star, both chiral sectors have the mean value and the variance of the ingoing sector of the Unruh state (Appendix~\ref{app:B}). Far from the source the relative fluctuation is
\begin{equation}\label{eq:RBinf}
\mathcal R_B(r)\simeq\frac{1}{c}\Big[\frac{22}{5}+a\Big(\frac{18}{5}\frac{r}{M}-3\Big)\Big],\qquad r\gg M .
\end{equation}
For $a\neq0$ it grows linearly with the distance, because the variance, $a\,cM/160\pi^2r^5$, is of first order in $M$, whereas $\langle T_{uu}\rangle^2$ is of second order. It remains finite, and equal to $22/5c$, only for $a=0$.

The dependence on $\alpha$ can be summarized as follows (see Table~\ref{tab:alpha} in Appendix~\ref{app:B}). The relative fluctuation of the Hawking flux at infinity is $22/5c$ for every $\alpha$. In the exterior region the outgoing and the ingoing variances are both non-negative if and only if $-3/10\leq\alpha\leq0$. The standard normal-ordered product of conformal field theory \cite{DiFrancesco} is its upper edge and the quasi-primary field $\Lambda$ is the lower one. Only for the latter do the relative fluctuations of the outgoing and of the ingoing fluxes at the horizon coincide, and their common value is then the one at infinity.

\textbf{Conclusions}.---Two-dimensional conformal symmetry has had a broad impact in many areas of physics. In this paper we have used it to obtain, exactly and in closed form, the local fluctuations of the stress tensor of a conformal field theory in any two-dimensional background and in any quantum state, Eq.~(\ref{eq:variance}). This result extends Eq.~(\ref{eq:Tmean}) for the mean value $\langle T_{\pm\pm}\rangle$, which has played a central role in the study of quantum black holes \cite{CGHS92,Hawking92,Strominger95,FabbriNavarroSalas, APR11,HSS20,PSS23,AKPPS25}. The square of the stress tensor has been represented by the normal-ordered product $T^2$ and, more generally, by $T^2+\alpha\,\partial^2T$, where $\alpha$ labels different point-splitting prescriptions and enters the variance through a single term, proportional to the second covariant derivative of the mean value. The relative fluctuation scales as $1/c$ for any $\alpha$, which justifies the semiclassical approximation for a large number of matter fields. For the Schwarzschild black hole in the Unruh state, the Hawking flux at infinity has the relative fluctuation $22/5c$ of thermal radiation, independently of $\alpha$. At the horizon the relative fluctuation is $-44\alpha/3c$ for the outgoing flux and $(238+720\alpha)/5c$ for the ingoing flux, which controls the decrease of the black hole mass. Since Eq.~(\ref{eq:variance}) holds for any background and any state, these results have many other potential applications, which we leave for future work.

\textbf{Acknowledgments}.---This work has been supported by Project No.~PID2023-149560NB-C21 funded by MCIU/AEI/10.13039/501100011033/FEDER, UE, and by the Severo Ochoa Excellence Grant CEX2023-001292-S. The paper is based upon work from COST Action CaLISTA CA21109 supported by COST (European Cooperation in Science and Technology). Some computations were performed and checked with the help of Mathematica and of the AI tool Claude Fable 5.2 (Anthropic). The authors take full responsibility for their validity.

\clearpage
\onecolumngrid
\appendix
\setcounter{secnumdepth}{1}%
\setcounter{equation}{0}

\section{Normal ordering, point-splitting prescriptions and transformation laws}\label{app:A}

In this appendix we collect the properties of the operator product expansion and of the normal-ordered product that are used in the main text. We explain the origin of the parameter $\alpha$, and we derive the finite transformation law (\ref{eq:TTfinite}) of the operators $T^2_\alpha=T^2+\alpha\,\partial^2T$ under a holomorphic map $\omega=f(z)$. We use the notation of the main text, $p=f'$, $\sigma=f''/f'$, $S=\sigma'-\sigma^2/2$ and $a=3+10\alpha$, with primes denoting derivatives with respect to $z$.

\textit{Normal-ordered product.} Following Ref.~\cite{DiFrancesco}, the operator product of two fields is expanded in powers of $z-w$ with coefficients evaluated at the point $w$ of the second field, $A(z)B(w)=\sum_n\{AB\}_n(w)/(z-w)^n$, and the normal-ordered product is the term of order zero, $(AB)(w)=\{AB\}_0(w)$. For the stress tensor, $\{TT\}_4=c/2$, $\{TT\}_2=2T$ and $\{TT\}_1=\partial T$, Eq.~(\ref{eq:TTOPE}), and
\begin{equation}\label{eq:A-normal}
T^2(w)\equiv(TT)(w)=\lim_{z\to w}\Big[T(z)T(w)-\frac{c/2}{(z-w)^4}-\frac{2T(w)}{(z-w)^2}-\frac{\partial T(w)}{z-w}\Big]=\frac{1}{2\pi i}\oint_w\frac{dz}{z-w}\,T(z)T(w) .
\end{equation}
The convention that the coefficients are evaluated at $w$ is part of the definition. With it there is no ambiguity, since $T^2$ is a definite operator, and its expectation value in any state is the finite number obtained from Eq.~(\ref{eq:A-normal}). The definition is symmetric in the two points. If the coefficients are evaluated at $z$, Eq.~(\ref{eq:TTOPE}) reads $T(z)T(w)\sim\frac{c/2}{(z-w)^4}+\frac{2T(z)}{(z-w)^2}-\frac{\partial T(z)}{z-w}$, and the regular term is again $T^2$, as one checks by expanding $T(z)$ and $\partial T(z)$ around $w$.

\textit{Other point-splitting prescriptions.} Equation~(\ref{eq:A-normal}) is a point-splitting prescription, in which the two operators are separated, the singular terms are subtracted, and the points are brought together. A different prescription is obtained if the expansion is referred to an intermediate point $x_\lambda=w+\lambda(z-w)$, with $0\leq\lambda\leq1$. Expanding $\{AB\}_n(w)$ around $x_\lambda$,
\begin{equation}\label{eq:A-lambdaOPE}
A(z)B(w)=\sum_n\frac{\{AB\}^\lambda_n(x_\lambda)}{(z-w)^n},\qquad
\{AB\}^\lambda_n=\sum_{k\geq0}\frac{(-\lambda)^k}{k!}\,\partial^k\{AB\}_{n+k} ,
\end{equation}
and the term of order zero defines
\begin{equation}\label{eq:A-ABlambda}
(AB)_\lambda\equiv\{AB\}^\lambda_0=(AB)+\sum_{k>0}\frac{(-\lambda)^k}{k!}\,\partial^k\{AB\}_k .
\end{equation}
For $\lambda=0$ this is $(AB)$ and for $\lambda=1$ it is $(BA)$ \cite{DiFrancesco}, so that $(AB)_\lambda$ interpolates between the two orderings. Only the field-dependent singular terms of the operator product contribute to $(AB)_\lambda-(AB)$. For the stress tensor Eq.~(\ref{eq:A-lambdaOPE}) reads
\begin{equation}\label{eq:A-TTlambda}
T(z)T(w)=\frac{c/2}{(z-w)^4}+\frac{2T(x_\lambda)}{(z-w)^2}+\frac{(1-2\lambda)\,\partial T(x_\lambda)}{z-w}+(TT)_\lambda(x_\lambda)+O(z-w),\qquad
(TT)_\lambda=T^2-\lambda(1-\lambda)\,\partial^2T .
\end{equation}
This is the same expansion as Eq.~(\ref{eq:TTOPE}), written in terms of the operators at $x_\lambda$. Its singular terms, however, are not those of Eq.~(\ref{eq:TTOPE}), and if they are the ones that are subtracted, the finite remainder is $(TT)_\lambda$ and not $T^2$. The cases $\lambda=0$ and $\lambda=1$ coincide. Each $(TT)_\lambda$ is a well-defined operator. The parameter
\begin{equation}\label{eq:A-alpha}
\alpha=-\lambda(1-\lambda)
\end{equation}
therefore labels different prescriptions. The choices of the point span the interval $-1/4\leq\alpha\leq0$. Other values of $\alpha$ correspond to a further finite subtraction of $\partial^2T$, which is the only other operator of weight four that can be built from the stress tensor alone, and this is the reason for considering the family $T^2_\alpha$ with arbitrary $\alpha$. The differences between the prescriptions are limited. They are linear in the stress tensor, they do not affect the dependence on $c$ of the relative fluctuation, Eq.~(\ref{eq:ratio}), and they vanish in states in which $\langle T\rangle$ is constant, such as the thermal state of Appendix~\ref{app:T}.

\textit{Free fields.} The prescriptions coincide, $(AB)_\lambda=(AB)$, when the only singular term of the operator product of $A$ and $B$ has a constant coefficient, as for the elementary fields of a free theory. This is not the case of the stress tensor, even in a free theory, since Eq.~(\ref{eq:TTOPE}) has field-dependent singular terms for any $c$. For a free boson, with $J=\partial\varphi$, $J(z)J(w)\sim-1/(z-w)^2$ and $T=-\frac12{:}J^2{:}$, Wick's theorem gives
\begin{equation}\label{eq:A-wick}
T(z)T(w)=\frac{1/2}{(z-w)^4}-\frac{1}{(z-w)^2}\,{:}J(z)J(w){:}+\frac14\,{:}J^2(z)J^2(w){:} .
\end{equation}
The last term is regular, and its coincidence limit, $\frac14{:}J^4{:}$, which is quartic in the field, is the Wick-ordered square of the stress tensor considered in Appendix~\ref{app:W}. It does not depend on the prescription. The term with a single contraction is a singular function multiplied by a regular bilocal operator, and its finite part depends on the point around which the bilocal operator is expanded. One finds
\begin{equation}\label{eq:A-TTlambdafree}
(TT)_\lambda=\frac14{:}J^4{:}-\frac{(1-\lambda)^2+\lambda^2}{2}\,{:}J''J{:}+\lambda(1-\lambda)\,{:}J'J'{:} ,
\end{equation}
in agreement with Eq.~(\ref{eq:A-TTlambda}), since $\partial^2T=-{:}J''J{:}-{:}J'J'{:}$. In a free theory the dependence on the prescription thus resides entirely in the term with a single contraction.

\textit{Operator product with $T$ and infinitesimal transformation.} Under $f(z)=z+\epsilon(z)$ the variation of a local operator is given by the conformal Ward identity \cite{DiFrancesco},
\begin{equation}\label{eq:A-ward}
\delta_\epsilon\mathcal O(w)=-\frac{1}{2\pi i}\oint_w dz\,\epsilon(z)\,T(z)\,\mathcal O(w),
\end{equation}
so that a term $\mathcal O_n(w)/(z-w)^n$ of the operator product of $T(z)$ with $\mathcal O(w)$ contributes $-\epsilon^{(n-1)}\mathcal O_n/(n-1)!$. Only the singular terms are needed. Two different operator products should be distinguished here. The singular terms of $T(z)T(w)$ are those subtracted in the definition of the composite operator. The transformation law of the composite operator is determined instead by the singular terms of its own operator product with $T(z)$, and these depend on which operator it is. Differentiating Eq.~(\ref{eq:TTOPE}) twice with respect to $w$,
\begin{equation}\label{eq:A-Td2T}
T(z)\,\partial^2T(w)\sim\frac{10c}{(z-w)^6}+\frac{12\,T(w)}{(z-w)^4}+\frac{10\,\partial T(w)}{(z-w)^3}+\frac{4\,\partial^2T(w)}{(z-w)^2}+\frac{\partial^3T(w)}{z-w},
\end{equation}
and together with Eq.~(\ref{eq:TTTOPE}) this gives
\begin{equation}\label{eq:A-TTaOPE}
T(z)\,T^2_\alpha(w)\sim\frac{c\,a}{(z-w)^6}+\frac{(c+8+12\alpha)\,T(w)}{(z-w)^4}+\frac{a\,\partial T(w)}{(z-w)^3}+\frac{4\,T^2_\alpha(w)}{(z-w)^2}+\frac{\partial T^2_\alpha(w)}{z-w}.
\end{equation}
The poles of order two and one are those of any operator of weight four. The other three depend on $\alpha$, and so does the variation,
\begin{equation}\label{eq:A-deltaTTa}
\delta_\epsilon T^2_\alpha=-\epsilon\,\partial T^2_\alpha-4\epsilon'\,T^2_\alpha-\frac{a}{2}\,\epsilon''\,\partial T-\frac{c+8+12\alpha}{6}\,\epsilon'''\,T-\frac{c\,a}{120}\,\epsilon^{(5)} .
\end{equation}
For the stress tensor itself Eqs.~(\ref{eq:A-ward}) and (\ref{eq:TTOPE}) give $\delta_\epsilon T=-\epsilon\,\partial T-2\epsilon'T-\frac{c}{12}\epsilon'''$, which is the infinitesimal form of Eq.~(\ref{eq:Tfinite}).

\textit{The operator $\Lambda$.} The poles of order six and three of Eq.~(\ref{eq:A-TTaOPE}) are absent for $a=0$, and only then. This is the case of
\begin{equation}\label{eq:Lambda}
\Lambda\equiv T^2-\frac{3}{10}\,\partial^2T ,\qquad T^2_\alpha=\Lambda+\frac{a}{10}\,\partial^2T ,
\end{equation}
for which
\begin{equation}\label{eq:TLambdaOPE}
T(z)\Lambda(w)\sim\frac{(c+22/5)\,T(w)}{(z-w)^4}+\frac{4\Lambda(w)}{(z-w)^2}+\frac{\partial\Lambda(w)}{z-w},\qquad
\delta_\epsilon\Lambda=-\epsilon\,\partial\Lambda-4\epsilon'\Lambda-\frac{5c+22}{30}\,\epsilon'''\,T .
\end{equation}
The absence of the pole of order three is the statement that $\Lambda$ is quasi-primary \cite{Zamolodchikov85}, and the absence of the pole of order six is the statement that $\langle\Lambda(z)T(w)\rangle=0$ in the vacuum. The value $\alpha=-3/10$ is not obtained from Eq.~(\ref{eq:A-alpha}) for any real $\lambda$, and $\Lambda$ is the normal-ordered product minus a definite multiple of $\partial^2T$.

\textit{Finite transformation of $\Lambda$.} The finite law must be local in $f$, of total weight four, and it must reduce to the variation in Eq.~(\ref{eq:TLambdaOPE}) at first order in $\epsilon$, where $p\simeq1+\epsilon'$, $\sigma\simeq\epsilon''$ and $S\simeq\epsilon'''$. Since that variation contains neither a term in $\epsilon''\partial T$ nor a c-number term, we look for a law in which the map enters only through the Schwarzian derivative,
\begin{equation}\label{eq:A-Lambdafinite}
\widetilde\Lambda(\omega)=p^{-4}\big[\Lambda(z)+\mathcal A\,S\,T(z)+\mathcal C\,S^2\big],\qquad\mathcal A=-\frac{5c+22}{30}.
\end{equation}
The term $\mathcal C\,S^2$ is of second order in $\epsilon$, and the coefficient $\mathcal C$ is fixed by the consistency of Eq.~(\ref{eq:A-Lambdafinite}) under the composition of two maps, $z\to\omega=f(z)\to\eta=g(\omega)$, with $p=f'(z)$ and $q=g'(\omega)$, so that the derivative of the composed map is $pq$. The Schwarzian derivative obeys the cocycle property $S_{g\circ f}(z)=S_f(z)+p^2\,S_g(\omega)$. Applying Eq.~(\ref{eq:A-Lambdafinite}) twice, and using Eq.~(\ref{eq:Tfinite}) for the intermediate $\widetilde T(\omega)=p^{-2}[T(z)-\frac{c}{12}S_f]$, we obtain
\begin{equation}
(pq)^{-4}\Big[\Lambda+\mathcal A\,S_fT+\mathcal C\,S_f^2+\mathcal A\,p^2S_g\Big(T-\frac{c}{12}S_f\Big)+\mathcal C\,p^4S_g^2\Big],
\end{equation}
whereas a single application with $S_{g\circ f}$ gives
\begin{equation}
(pq)^{-4}\Big[\Lambda+\mathcal A\,\big(S_f+p^2S_g\big)T+\mathcal C\,\big(S_f+p^2S_g\big)^2\Big].
\end{equation}
The terms in $T$, $S_f^2$ and $S_g^2$ coincide, and the mixed terms $S_fS_g$ coincide only if $2\mathcal C=-\mathcal A c/12$. Hence
\begin{equation}\label{eq:Lambdafinite}
\widetilde\Lambda(\omega)=p^{-4}\Big[\Lambda-\frac{5c+22}{30}\,S\,T+\frac{c\,(5c+22)}{720}\,S^2\Big].
\end{equation}

\textit{Finite transformation of $T^2_\alpha$.} From Eq.~(\ref{eq:Lambda}), $\widetilde{T^2_\alpha}=\widetilde\Lambda+\frac{a}{10}\partial_\omega^2\widetilde T$. With $\partial_\omega=p^{-1}\partial_z$ and Eq.~(\ref{eq:Tfinite}),
\begin{equation}\label{eq:A-d2T}
\partial_\omega^2\widetilde T=p^{-4}\Big[X''-5\sigma X'+\big(6\sigma^2-2\sigma'\big)X\Big],\qquad X=T-\frac{c}{12}S,
\end{equation}
and $6\sigma^2-2\sigma'=5\sigma^2-2S$. Collecting terms, the coefficient of $S\,T$ is $\mathcal A-\frac{a}{5}=-\frac{c+8+12\alpha}{6}$, and that of $S^2$ is $\mathcal C+\frac{c\,a}{60}=\frac{c(5c+58+120\alpha)}{720}$. The result is Eq.~(\ref{eq:TTfinite}) of the main text,
\begin{equation}\label{eq:A-TTfinite}
\widetilde{T^2_\alpha}(\omega)=p^{-4}\bigg[T^2_\alpha-\frac{a}{2}\sigma\,\partial T+\frac{a}{2}\sigma^2T-\frac{c+8+12\alpha}{6}S\,T-\frac{c\,a}{120}S''+\frac{c\,a}{24}\sigma S'-\frac{c\,a}{24}\sigma^2S+\frac{c(5c+58+120\alpha)}{720}S^2\bigg].
\end{equation}
For the normal-ordered product, $\alpha=0$ and $a=3$, it reads
\begin{equation}\label{eq:A-TTfinite0}
\widetilde{T^2}(\omega)=p^{-4}\bigg[T^2-\frac{3}{2}\sigma\,\partial T+\frac{9\sigma^2-(c+8)S}{6}\,T+\frac{c}{720}\big[(5c+58)S^2+90\sigma(S'-\sigma S)-18S''\big]\bigg].
\end{equation}
For a map with $S=0$ identically, a M\"obius map $\omega=(Az+B)/(Cz+D)$, Eq.~(\ref{eq:A-TTfinite}) reduces to $\widetilde{T^2_\alpha}=p^{-4}[T^2_\alpha-\frac{a}{2}\sigma(\partial T-\sigma T)]$, with $\sigma=-2C/(Cz+D)$. For any $\alpha$ the operator $T^2_\alpha$ is a tensor under the linear maps, $C=0$, which in Lorentzian signature are the translations, the Lorentz boosts and the dilatations of the null coordinates. Only for $a=0$ is it a tensor under all the M\"obius maps, as the stress tensor is. Therefore, the value $a=0$, that is, $\alpha=-3/10$, is distinguished. In this case $T^2_\alpha=\Lambda$, and the fluctuations are covariant under all M\"obius transformations.

\textit{Checks.} (i) In the infinitesimal limit Eqs.~(\ref{eq:Lambdafinite}) and (\ref{eq:A-TTfinite}) reproduce the variations in Eqs.~(\ref{eq:TLambdaOPE}) and (\ref{eq:A-deltaTTa}). (ii) They are consistent under composition, since the law of $\Lambda$ is consistent by construction and $\partial^2T$ transforms consistently by Eq.~(\ref{eq:Tfinite}).

\textit{Vacuum states.} Consider the vacuum $|0_\omega\rangle$ defined by the coordinate $\omega$, in which $\langle\widetilde T\rangle=\langle\widetilde\Lambda\rangle=\langle\widetilde{T^2_\alpha}\rangle=0$. Then Eq.~(\ref{eq:Tfinite}) gives $\langle T\rangle=\frac{c}{12}S$, Eq.~(\ref{eq:Lambdafinite}) gives $\langle\Lambda\rangle=\frac{c(5c+22)}{720}S^2$, and Eq.~(\ref{eq:A-TTfinite}) gives $\langle T^2_\alpha\rangle=\frac{c\,a}{120}S''+\frac{c(5c+22)}{720}S^2$, after the cancellation of all the terms that contain $\sigma$. The only term quadratic in $c$, $c^2S^2/144$, is the square of the anomalous term of Eq.~(\ref{eq:Tfinite}), so that it cancels in the variance,
\begin{equation}\label{eq:A-var}
\langle T^2_\alpha\rangle-\langle T\rangle^2=\frac{c\,a}{120}\,S''+\frac{11c}{360}\,S^2 ,\qquad
\langle\Lambda\rangle-\langle T\rangle^2=\frac{11c}{360}\,S^2=\frac{22}{5c}\,\langle T\rangle^2 .
\end{equation}

\section{Variance in curved space and in the Schwarzschild geometry}\label{app:B}

In this appendix we give the intermediate steps that lead to Eqs.~(\ref{eq:variance}) and (\ref{eq:covvac}) of the main text, and some additional results for the Schwarzschild black hole.

\textit{Locally inertial coordinates.} The expansion of Eq.~(\ref{eq:xiX}) around $X$, with $\Delta=x^\pm-x^\pm(X)$ and $\rho_n=\partial_\pm^n\rho(X)$, is
\begin{equation}\label{eq:xiexpansion}
\frac{\xi_X^\pm}{b_\pm}=\Delta+\rho_1\Delta^2+\frac{\rho_2+2\rho_1^2}{3}\,\Delta^3+\frac{\rho_3+6\rho_1\rho_2+4\rho_1^3}{12}\,\Delta^4+\frac{\rho_4+8\rho_1\rho_3+6\rho_2^2+24\rho_1^2\rho_2+8\rho_1^4}{60}\,\Delta^5+O(\Delta^6),
\end{equation}
which coincides with the purely chiral part of the Riemann normal coordinates at $X$. The choice in Eq.~(\ref{eq:xiX}) is unique up to the local boost $b_\pm\to e^{\pm\beta}b_\pm$, which does not affect the results. Along each null line through $X$ the conformal factor in the new coordinates is constant, so that $\xi_X^\pm$ is locally inertial, in the corresponding chiral sector, at every point of the line. Since Eq.~(\ref{eq:xiX}) holds along the whole null line, $\sigma_\pm=2\partial_\pm\rho$ there, and at $X$
\begin{equation}\label{eq:sigmaS}
\sigma_\pm=2\rho_1,\qquad S_\pm=2\rho_2-2\rho_1^2,\qquad
S_\pm'=2\rho_3-4\rho_1\rho_2,\qquad S_\pm''=2\rho_4-4\rho_1\rho_3-4\rho_2^2 .
\end{equation}
For the stress tensor and for $\Lambda$, whose transformation laws involve only the Schwarzian derivative, the expansion up to third order is sufficient. For $a\neq0$ Eq.~(\ref{eq:TTfinite}) involves $S''$, and the expansion is needed up to fifth order.

\textit{Second moment and variance.} In physical normalization Eq.~(\ref{eq:Lambda}) reads ${:}T_{\pm\pm}^2{:}_\alpha={:}\Lambda_{\pm\pm}{:}-\frac{a}{20\pi}\partial_\pm^2{:}T_{\pm\pm}{:}$. Consider first $\Lambda$. The prescription (\ref{eq:prescription}) and Eq.~(\ref{eq:Lambdafinite}), with $z=x^\pm$ and $\omega=\xi_X^\pm$, give
\begin{equation}\label{eq:T2mean}
\langle\Lambda_{\pm\pm}\rangle=\langle{:}\Lambda_{\pm\pm}{:}\rangle+\frac{(5c+22)\,S_\pm}{60\pi}\,\langle{:}T_{\pm\pm}{:}\rangle+\frac{c\,(5c+22)}{2880\pi^2}\,S_\pm^2 .
\end{equation}
For general $\alpha$ there is the additional term $-\frac{a}{20\pi}\,p_\pm^4\,\partial_\xi^2\langle{:}T_{\pm\pm}(\xi_X^\pm){:}\rangle|_X$. Since the coordinates (\ref{eq:xiX}) are locally inertial along the null line through $X$, $\langle{:}T_{\pm\pm}(\xi_X^\pm){:}\rangle$ is the physical expectation value on that line, and its second derivative at $X$ is the second covariant derivative. Hence
\begin{equation}\label{eq:T2alpha}
\langle T_{\pm\pm}^2\rangle_\alpha=\langle\Lambda_{\pm\pm}\rangle-\frac{a}{20\pi}\,\nabla_\pm\nabla_\pm\langle T_{\pm\pm}\rangle ,\qquad
\nabla_\pm\nabla_\pm\langle T_{\pm\pm}\rangle=\big[\partial_\pm^2-10\rho_1\partial_\pm+24\rho_1^2-4\rho_2\big]\langle T_{\pm\pm}\rangle .
\end{equation}
To obtain the variance we subtract $\langle T_{\pm\pm}\rangle^2=\langle{:}T_{\pm\pm}{:}\rangle^2+(cS_\pm/12\pi)\langle{:}T_{\pm\pm}{:}\rangle+c^2S_\pm^2/576\pi^2$. The last term cancels the term quadratic in $c$ in Eq.~(\ref{eq:T2mean}), and the terms in $S_\pm\langle{:}T_{\pm\pm}{:}\rangle$ combine into $(5c+22)/60\pi-c/12\pi=11/30\pi$, independent of $c$. With Eq.~(\ref{eq:sigmaS}) this is Eq.~(\ref{eq:variance}). Using Eq.~(\ref{eq:Tmean}), it can also be written as
\begin{equation}\label{eq:covariant}
\langle T_{\pm\pm}^2\rangle_\alpha-\langle T_{\pm\pm}\rangle^2=\langle{:}\Lambda_{\pm\pm}{:}\rangle-\Big(1+\frac{22}{5c}\Big)\langle{:}T_{\pm\pm}{:}\rangle^2+\frac{22}{5c}\,\langle T_{\pm\pm}\rangle^2-\frac{a}{20\pi}\,\nabla_\pm\nabla_\pm\langle T_{\pm\pm}\rangle .
\end{equation}
The same result follows directly from Eq.~(\ref{eq:TTfinite}) and Eq.~(\ref{eq:sigmaS}). In terms of $\langle{:}T_{\pm\pm}^2{:}_\alpha\rangle$ and of the derivatives of $\rho$,
\begin{align}\label{eq:A-variance}
\langle T_{\pm\pm}^2\rangle_\alpha-\langle T_{\pm\pm}\rangle^2={}&\langle{:}T_{\pm\pm}^2{:}_\alpha\rangle-\langle{:}T_{\pm\pm}{:}\rangle^2+\frac{a\,\rho_1}{2\pi}\,\partial_\pm\langle{:}T_{\pm\pm}{:}\rangle+\frac{11(\rho_2-\rho_1^2)+3a(\rho_2-6\rho_1^2)}{15\pi}\,\langle{:}T_{\pm\pm}{:}\rangle\notag\\
&+\frac{c}{720\pi^2}\Big[22(\rho_2-\rho_1^2)^2+3a\big(24\rho_1^4-48\rho_1^2\rho_2+6\rho_2^2+12\rho_1\rho_3-\rho_4\big)\Big].
\end{align}
For the normal-ordered product, $\alpha=0$, the coefficient of $\langle{:}T_{\pm\pm}{:}\rangle$ is $(4\rho_2-13\rho_1^2)/3\pi$ and the geometric term is the one quoted in the main text, which to first order in the geometry reduces to $-\frac{c}{80\pi^2}\partial_\pm^4\rho$ and has no definite sign.

\textit{Vacuum states.} Consider the vacuum state associated with an arbitrary choice of null coordinates $y^\pm(x^\pm)$. In such a state $\langle{:}T_{\pm\pm}(y^\pm){:}\rangle=\langle{:}\Lambda_{\pm\pm}(y^\pm){:}\rangle=0$, and Eqs.~(\ref{eq:Tfinite}) and (\ref{eq:Lambdafinite}) give $\langle{:}T_{\pm\pm}(x^\pm){:}\rangle=-\frac{c}{24\pi}S_y$ and $\langle{:}\Lambda_{\pm\pm}(x^\pm){:}\rangle=\frac{c(5c+22)}{2880\pi^2}S_y^2$, with $S_y$ the Schwarzian derivative of the map $x^\pm\to y^\pm$. The first two terms of Eq.~(\ref{eq:covariant}) then cancel, which gives Eq.~(\ref{eq:covvac}). For $a=0$,
\begin{equation}\label{eq:universal}
\langle\Lambda_{\pm\pm}\rangle-\langle T_{\pm\pm}\rangle^2=\frac{22}{5c}\,\langle T_{\pm\pm}\rangle^2 .
\end{equation}
The reason is that $T$ and $\Lambda$ are sensitive to the map between $y^\pm$ and the locally inertial coordinate only through its Schwarzian derivative at the point, which is a single number. Any map agrees at a point, up to third order, with a map of vanishing Schwarzian derivative composed with the exponential $e^{\kappa\xi}$, whose Schwarzian derivative is $-\kappa^2/2$. To the order that $T$ and $\Lambda$ can detect, every vacuum state is therefore locally indistinguishable from a thermal state with an effective temperature $T_{\rm eff}$ fixed by $\langle T_{\pm\pm}\rangle=\frac{\pi c}{12}T_{\rm eff}^2$, and $22/5c$ is the relative fluctuation of thermal radiation in flat space (Appendix~\ref{app:T}). For $a\neq0$ the relative fluctuation depends on the point through $\nabla_\pm\nabla_\pm\langle T_{\pm\pm}\rangle/\langle T_{\pm\pm}\rangle^2$, which is sensitive to the variation of the effective temperature along the null line. For the vacuum associated with the coordinates $x^\pm$ in a weakly curved geometry, Eq.~(\ref{eq:covvac}) gives, to first order in $\rho$, the variance $-\frac{a\,c}{240\pi^2}\partial_\pm^4\rho$, while $\langle T_{\pm\pm}\rangle^2$ is of second order.

\textit{Outgoing flux in the Unruh state.} From $\langle U|{:}\Lambda_{UU}{:}|U\rangle=0$, Eq.~(\ref{eq:Lambdafinite}) gives $\langle U|{:}\Lambda_{uu}{:}|U\rangle=(c^2+22c/5)(\kappa^2/48\pi)^2$, and Eq.~(\ref{eq:varianceU}) follows from Eq.~(\ref{eq:variance}), or directly from Eq.~(\ref{eq:covvac}). The relative fluctuation, which does not depend on the coordinates, is
\begin{equation}\label{eq:Rout}
\mathcal R_{\rm out}(r)=\frac{2}{5c}\,\frac{N_\alpha(r)}{(r^2+4Mr+12M^2)^2},\qquad\mathcal R_{\rm out}(2M)=-\frac{44\,\alpha}{3c},\qquad\mathcal R_{\rm out}(\infty)=\frac{22}{5c}.
\end{equation}
The value at infinity does not depend on $\alpha$ because $\nabla_u\nabla_u\langle U|T_{uu}|U\rangle$ vanishes there. Equation~(\ref{eq:0variance}) is $(du/dU)^4$ times Eq.~(\ref{eq:varianceU}), evaluated at $r=2M$. These results are not simply a consequence of the Unruh state being the vacuum associated with $U$. The coordinate $U$ is regular but not locally inertial at the future horizon, where $\Gamma^U_{UU}=e^{\kappa v-1}/2M\neq0$, and in the frame $(U,v)$ both the mean value and the variance are given entirely by the geometric terms of Eqs.~(\ref{eq:Tmean}) and (\ref{eq:variance}).

\textit{Ingoing flux.} Changing coordinates $(u,v)\to(v,r)$ one gets $T_{vv}^{EF}=T_{uu}+T_{vv}+2T_{uv}$, since $u=v-2r^*$ and the vector $\partial_v$ at fixed $r$ is $\partial_v|_u+\partial_u|_v$, the timelike Killing vector. The three contributions behave differently. The trace is a c-number fixed by the geometry, $T_{uv}=-\frac{c}{12\pi}\partial_u\partial_v\rho=-\frac{cM(r-2M)}{24\pi r^4}$, and its connected correlators with $T_{uu}$ and $T_{vv}$ vanish. The operator product of $T_{uu}$ and $T_{vv}$, which belong to different chiral sectors, is regular, and the Unruh state is the product of the vacuum associated with $U$ in the outgoing sector and of the vacuum associated with $v$ in the ingoing sector. Therefore $\langle T_{uu}T_{vv}\rangle-\langle T_{uu}\rangle\langle T_{vv}\rangle=0$. The only composite operators that need to be defined are the two chiral squares, for which we take $T^2_\alpha$ with the same $\alpha$, and the variance is the sum (\ref{eq:varEF}). The first term is Eq.~(\ref{eq:varianceU}). For the second, $\langle U|{:}T_{vv}{:}|U\rangle=0$ and $\langle U|T_{vv}|U\rangle=-cM(2r-3M)/48\pi r^4$ \cite{FabbriNavarroSalas}, and Eq.~(\ref{eq:covvac}) gives
\begin{equation}\label{eq:varin}
\langle U|T_{vv}^2|U\rangle_\alpha-\langle U|T_{vv}|U\rangle^2=\frac{cM}{5760\,\pi^2r^8}\Big[108r^3-370Mr^2+480M^2r-225M^3+60\,\alpha\big(6r^3-23Mr^2+34M^2r-18M^3\big)\Big].
\end{equation}
On the horizon it gives
\begin{equation}\label{eq:varinH}
\langle (T_{vv}^{EF})^2\rangle_\alpha-\langle T_{vv}^{EF}\rangle^2\big|_{r=2M}=\frac{(119+360\,\alpha)\,c\kappa^4}{5760\pi^2}.
\end{equation}

\textit{Boulware state.} In this state $\langle B|{:}T_{uu}{:}|B\rangle=\langle B|{:}T_{uu}^2{:}_\alpha|B\rangle=0$, and likewise for $v$. The relative fluctuation, for any $r$, is
\begin{equation}\label{eq:RB}
\mathcal R_B(r)=\frac{2}{5c}\,\frac{108r^3-370Mr^2+480M^2r-225M^3+60\,\alpha\big(6r^3-23Mr^2+34M^2r-18M^3\big)}{M(2r-3M)^2},
\end{equation}
of which Eq.~(\ref{eq:RBinf}) is the leading behavior. For $\alpha=-3/10$ the numerator is $11M(2r-3M)^2$ and $\mathcal R_B=22/5c$. At infinity the Boulware state approaches the Minkowski vacuum, in which the mean value and the renormalized fluctuations of the stress tensor vanish. For $a\neq0$ the relative fluctuation nevertheless grows without bound, since $\langle T_{uu}\rangle^2\simeq c^2M^2/576\pi^2r^6$.

\textit{Summary of the dependence on $\alpha$.} Table~\ref{tab:alpha} collects the relative fluctuations for several values of $\alpha$. In the exterior region $r\geq2M$ the outgoing variance (\ref{eq:varianceU}) is non-negative if and only if $\alpha\leq0$, and the ingoing variance (\ref{eq:varin}) if and only if $\alpha\geq-3/10$.

\begin{table}[h]
\caption{\label{tab:alpha}Relative fluctuation, multiplied by $c$, of the outgoing flux at the horizon ($\mathcal R_{\rm out}^{H}$), of the ingoing flux at the horizon ($\mathcal R_{\rm in}^{H}$), and of the Hawking flux at infinity ($\mathcal R^{\infty}$), in the Unruh state, and leading behavior for $r\gg M$ in the Boulware state ($\mathcal R_B$), for the normal-ordered product ($\alpha=0$), the quasi-primary field $\Lambda$ ($\alpha=-3/10$), and general $\alpha$, with $a=3+10\alpha$.}
\begin{ruledtabular}
\begin{tabular}{lcccc}
$\alpha$ & $c\,\mathcal R_{\rm out}^{H}$ & $c\,\mathcal R_{\rm in}^{H}$ & $c\,\mathcal R^{\infty}$ & $c\,\mathcal R_B$\\
\hline
$0$ & $0$ & $238/5$ & $22/5$ & $54r/5M$\\
$-3/10$ & $22/5$ & $22/5$ & $22/5$ & $22/5$\\
any & $-44\alpha/3$ & $(238+720\alpha)/5$ & $22/5$ & $18a\,r/5M$\\
\end{tabular}
\end{ruledtabular}
\end{table}

\section{Fluctuations defined by ${:}\big(\sum_i(\partial_\pm\phi_i)^2\big)^2{:}$}\label{app:W}

For free fields the square of the stress tensor can also be represented by an operator built directly from the fields, as in Ref.~\cite{PerezSudarsky26}. [That reference considers a single scalar field. Here we extend the analysis to $c$ fields.] For $N=c$ massless scalar fields $\phi_i$, with $T_{\pm\pm}=\sum_i(\partial_\pm\phi_i)^2$ in the physical normalization of the main text, the natural operator is the square of the total stress tensor with all self-contractions removed,
\begin{equation}\label{eq:W-def}
{:}W_{\pm\pm}{:}\equiv{:}\Big(\sum_i(\partial_\pm\phi_i)^2\Big)^2{:}=\sum_i{:}(\partial_\pm\phi_i)^4{:}+\sum_{i\neq j}{:}(\partial_\pm\phi_i)^2{:}\,{:}(\partial_\pm\phi_j)^2{:} .
\end{equation}
The two sums have a different origin. For $i\neq j$ the fields are independent, there are no contractions between them, and the product ${:}(\partial_\pm\phi_i)^2{:}\,{:}(\partial_\pm\phi_j)^2{:}$ at the same point is finite. For $i=j$ the product of the two stress tensors diverges, and it is defined by normal ordering the four fields together.

\textit{Variance.} The field $\partial_\pm\phi$ is primary, of weight one, and the normal orderings with respect to $\xi^\pm$ and to $x^\pm$ differ only in the subtracted contraction, which in the coincidence limit amounts to $S_\pm/24\pi$. Wick's theorem then gives
\begin{equation}\label{eq:W-law}
p_\pm^4\,{:}W_{\pm\pm}(\xi^\pm){:}={:}W_{\pm\pm}{:}+\frac{(c+2)\,S_\pm}{12\pi}\,{:}T_{\pm\pm}{:}+\frac{c\,(c+2)\,S_\pm^2}{576\pi^2},
\end{equation}
which is the transformation law (\ref{eq:Lambdafinite}) of $\Lambda$, in physical normalization, with $22/5$ replaced by $2$. With the prescription (\ref{eq:prescription}), and proceeding as in Appendix~\ref{app:B},
\begin{equation}\label{eq:W-var}
\langle W_{\pm\pm}\rangle-\langle T_{\pm\pm}\rangle^2=\langle{:}W_{\pm\pm}{:}\rangle-\Big(1+\frac{2}{c}\Big)\langle{:}T_{\pm\pm}{:}\rangle^2+\frac{2}{c}\,\langle T_{\pm\pm}\rangle^2 ,
\end{equation}
to be compared with Eq.~(\ref{eq:covariant}) for $a=0$. In the vacuum state associated with any choice of null coordinates the fields are Gaussian, with vanishing one-point function, so that $\langle{:}(\partial_\pm\phi_i)^4{:}\rangle=3\langle{:}(\partial_\pm\phi_i)^2{:}\rangle^2$. If all the fields are in the same state, $\langle{:}W_{\pm\pm}{:}\rangle=[3c+c(c-1)]\langle{:}T_{\pm\pm}{:}\rangle_{(1)}^2=(1+\frac2c)\langle{:}T_{\pm\pm}{:}\rangle^2$, the first two terms of Eq.~(\ref{eq:W-var}) cancel, and
\begin{equation}\label{eq:W-vac}
\langle W_{\pm\pm}\rangle-\langle T_{\pm\pm}\rangle^2=\frac{2}{c}\,\langle T_{\pm\pm}\rangle^2
\end{equation}
at every point and in any background. The terms quadratic in $c$ cancel, as in Eq.~(\ref{eq:variance}), and the relative fluctuation is $2/c$. For $c=1$ this is the value $2$ found in Ref.~\cite{PerezSudarsky26}.
Therefore, the validity of the semiclassical approximation is guaranteed when $c\gg1$. This differs from the conclusion of Ref.~\cite{PerezSudarsky26}.

\textit{Comparison with $T^2_\alpha$.} Wick's theorem applied to the product $T(z)T(w)$, Eq.~(\ref{eq:A-wick}), gives
\begin{equation}\label{eq:W-rel}
{:}\Lambda_{\pm\pm}{:}={:}W_{\pm\pm}{:}-\frac{1}{5\pi}\sum_i{:}(\partial_\pm^3\phi_i)(\partial_\pm\phi_i){:}+\frac{3}{10\pi}\sum_i{:}(\partial_\pm^2\phi_i)^2{:} ,
\end{equation}
and ${:}T_{\pm\pm}^2{:}_\alpha={:}\Lambda_{\pm\pm}{:}-\frac{a}{20\pi}\partial_\pm^2{:}T_{\pm\pm}{:}$. The operator $W_{\pm\pm}$ is therefore not a member of the family $T^2_\alpha$ for any $\alpha$. It differs from it by operators bilinear in the fields, which are not built from the stress tensor alone. The bilinear terms are the finite part of the term with a single contraction in Eq.~(\ref{eq:A-wick}). The Wick ordering of the four fields removes this term altogether, whereas the subtraction of the singular terms of the operator product removes only its poles. In the vacuum states its finite part amounts to $\frac{12}{5c}\langle T_{\pm\pm}\rangle^2$ for $\Lambda$, which is the difference between $22/5c$ and $2/c$.

\textit{Free fermions.} The term with a single contraction cannot be discarded in general. For $N=2c$ free Majorana fermions $\psi_i$ the stress tensor is $T_{\pm\pm}=\sum_iT^{(i)}_{\pm\pm}$, with $T^{(i)}_{\pm\pm}$ bilinear in $\psi_i$, and the same prescription gives $\sum_{i\neq j}{:}T^{(i)}_{\pm\pm}{:}\,{:}T^{(j)}_{\pm\pm}{:}$, since the terms with $i=j$ vanish by Fermi statistics. In the vacuum states the resulting variance is negative, $-\frac{1}{2c}\langle T_{\pm\pm}\rangle^2$. The operator ${:}T_{\pm\pm}^2{:}_\alpha$ gives instead the relative fluctuation $22/5c$ for $\alpha=-3/10$, as for any conformal field theory.

\section{Thermal state in flat space}\label{app:T}

In this appendix we obtain the local variance of thermal radiation in flat space directly from the two-point function of the stress tensor, and we write in physical normalization the point-splitting prescriptions of Appendix~\ref{app:A}. We use null coordinates $u=t-x$, $v=t+x$, with $ds^2=-du\,dv$, and a thermal state at temperature $\kappa/2\pi$. The two chiral sectors are independent, $\langle T_{uu}T_{vv}\rangle-\langle T_{uu}\rangle\langle T_{vv}\rangle=0$, and we consider the outgoing sector. The mean value is
\begin{equation}\label{eq:C-mean}
\langle T_{uu}\rangle=\frac{c\kappa^2}{48\pi}.
\end{equation}

\textit{Two-point function.} The connected two-point function of the outgoing flux is
\begin{equation}\label{eq:C-twopoint}
\langle T_{uu}(u)T_{uu}(u')\rangle-\langle T_{uu}\rangle^2=\frac{c\kappa^4}{8\pi^2}\,\frac{e^{2\kappa\Delta}}{(e^{\kappa\Delta}-1)^4}=\frac{c\kappa^4}{128\pi^2}\,\frac{1}{\sinh^4(\kappa\Delta/2)},\qquad \Delta=u-u'.
\end{equation}
The expression is even in $\Delta$, and we omit the usual prescription $\Delta\to\Delta-i\epsilon$, which does not affect the coincidence limit. For a single field this expression is given in Ref.~\cite{Wilczek93}, and for central charge $c$ it is multiplied by $c$. It can also be obtained from the vacuum correlator $c/(8\pi^2\Delta^4)$, since the thermal state is, in the outgoing sector, the vacuum associated with the coordinate $U=-\kappa^{-1}e^{-\kappa u}$. The anomalous c-number terms of Eq.~(\ref{eq:Tfinite}) cancel in the connected function, which becomes $c\,U'(u)^2U'(u')^2/[8\pi^2(U(u)-U(u'))^4]$, and this coincides with Eq.~(\ref{eq:C-twopoint}). At short distances
\begin{equation}\label{eq:C-expansion}
\langle T_{uu}(u)T_{uu}(u')\rangle-\langle T_{uu}\rangle^2=\frac{c}{8\pi^2\Delta^4}-\frac{c\kappa^2}{48\pi^2\Delta^2}+\frac{11c\kappa^4}{5760\pi^2}+O(\Delta^2).
\end{equation}

\textit{Local variance.} The two singular terms of Eq.~(\ref{eq:C-expansion}) are those of the operator product expansion (\ref{eq:TTOPE}). In physical normalization, ${:}T_{uu}{:}=-T/2\pi$, and including the regular term, $\TT=\Lambda+\frac{3}{10}\partial^2T$, Eq.~(\ref{eq:TTOPE}) reads
\begin{equation}\label{eq:C-OPE}
T_{uu}(u)T_{uu}(u')=\frac{c}{8\pi^2\Delta^4}-\frac{T_{uu}(u')}{\pi\Delta^2}-\frac{\partial_uT_{uu}(u')}{2\pi\Delta}-\frac{3}{20\pi}\,\partial_u^2T_{uu}(u')+{:}\Lambda_{uu}{:}(u')+O(\Delta).
\end{equation}
In the thermal state $\langle T_{uu}\rangle$ is constant, the terms with derivatives vanish, so that $\langle{:}T_{uu}^2{:}_\alpha\rangle=\langle{:}\Lambda_{uu}{:}\rangle$ for any $\alpha$, and the second term gives exactly $-c\kappa^2/48\pi^2\Delta^2$. The term in $\Delta^{-2}$ is therefore not a geometric singularity. It is the pole of order two of the operator product expansion, and its coefficient is proportional to the mean value of the stress tensor in the state. Hence
\begin{equation}\label{eq:C-local}
\langle T_{uu}^2\rangle_\alpha-\langle T_{uu}\rangle^2=\lim_{\Delta\to0}\Big[\langle T_{uu}(u)T_{uu}(u')\rangle-\langle T_{uu}\rangle^2-\frac{c}{8\pi^2\Delta^4}+\frac{c\kappa^2}{48\pi^2\Delta^2}\Big]=\frac{11c\kappa^4}{5760\pi^2}.
\end{equation}
This agrees with Eq.~(\ref{eq:variance}) for $\rho=0$, with the asymptotic value of Eq.~(\ref{eq:varianceU}), and with Eq.~(\ref{eq:A-var}) for $S=-\kappa^2/2$. The relative fluctuation is $22/5c$. If only the leading term $c/8\pi^2\Delta^4$ were subtracted, the coincidence limit would still diverge, with a coefficient proportional to $\langle T_{uu}\rangle$.

\textit{Comparison with the Wick-ordered square.} The result (\ref{eq:C-local}) does not depend on $\alpha$. The Wick-ordered square of Appendix~\ref{app:W} gives a different value. The thermal state is the vacuum state associated with the coordinate $U$, and Eq.~(\ref{eq:W-vac}) gives
\begin{equation}\label{eq:C-Wthermal}
\langle W_{uu}\rangle-\langle T_{uu}\rangle^2=\frac{2}{c}\langle T_{uu}\rangle^2=\frac{c\kappa^4}{1152\pi^2}.
\end{equation}
The origin of the difference is seen in terms of modes. For $c$ free scalar fields $\langle\partial_u\phi(u)\partial_u\phi(u')\rangle=-\frac{1}{4\pi\Delta^2}+D(\Delta)$ for each field, where the first term is the vacuum value and
\begin{equation}\label{eq:C-D}
D(\Delta)=\frac{1}{2\pi}\int_0^\infty d\omega\,\omega\,n(\omega)\cos\omega\Delta ,\qquad n(\omega)=\frac{1}{e^{2\pi\omega/\kappa}-1},
\end{equation}
is the thermal part, which is regular at $\Delta=0$ and linear in the occupation number, with $\langle T_{uu}\rangle=c\,D(0)$. The thermal state is Gaussian. For each field, Wick's theorem gives two equal contractions between the two points, and the connected two-point function of $(\partial_u\phi)^2$ is twice the square of $\langle\partial_u\phi(u)\partial_u\phi(u')\rangle$. Different fields are independent and do not contribute. Summing over the $c$ fields and expanding the square, the two-point function (\ref{eq:C-twopoint}) becomes the sum of three terms,
\begin{equation}\label{eq:C-wick}
\langle T_{uu}(u)T_{uu}(u')\rangle-\langle T_{uu}\rangle^2=2c\,\Big[-\frac{1}{4\pi\Delta^2}+D(\Delta)\Big]^2=\frac{c}{8\pi^2\Delta^4}-\frac{c\,D(\Delta)}{\pi\Delta^2}+2c\,D(\Delta)^2 .
\end{equation}
The first term is the state-independent pole. The last term is quadratic in the occupation number, and at $\Delta=0$ it is $2c\,D(0)^2=\frac{2}{c}\langle T_{uu}\rangle^2$, the value (\ref{eq:C-Wthermal}). The Wick ordering of the four fields keeps only this term. The cross term is linear in the occupation number. Its pole is $-\langle T_{uu}\rangle/\pi\Delta^2$, the one subtracted in Eq.~(\ref{eq:C-local}). Its finite part comes from the term of order $\Delta^2$ in the expansion of $D(\Delta)$, Eq.~(\ref{eq:C-D}). It is $\frac{c}{4\pi^2}\int_0^\infty d\omega\,\omega^3n(\omega)=\frac{c\kappa^4}{960\pi^2}$, and it is kept by the prescription (\ref{eq:C-local}) and discarded by the Wick ordering. The sum of the two finite contributions, $\frac{c\kappa^4}{1152\pi^2}+\frac{c\kappa^4}{960\pi^2}$, is the value (\ref{eq:C-local}).

\textit{Comparison with the thermodynamic fluctuation formula.} The traditional measure of the fluctuations of thermal radiation is the variance of the energy $E$ contained in a region of size $L\gg1/\kappa$. With $\langle E\rangle=L\langle T_{uu}\rangle=c\,\pi T^2L/12$, where $T=\kappa/2\pi$ denotes here the temperature, the canonical ensemble gives the thermodynamic fluctuation formula
\begin{equation}\label{eq:C-thermo}
\langle E^2\rangle-\langle E\rangle^2=T^2\,\frac{\partial\langle E\rangle}{\partial T}=\frac{c\kappa^3L}{48\pi^2},\qquad\qquad \frac{\langle E^2\rangle-\langle E\rangle^2}{\langle E\rangle^2}=\frac{48}{c\,\kappa L}.
\end{equation}
This result involves no renormalization, and it depends on the theory only through the central charge. It is $L$ times the integral over $\Delta$ of the two-point function (\ref{eq:C-twopoint}), to which the poles subtracted in Eq.~(\ref{eq:C-local}) do not contribute. The regular function in the bracket of Eq.~(\ref{eq:C-local}) has therefore the local variance as its value at $\Delta=0$ and $c\kappa^3/48\pi^2$ as its integral. Both variances are proportional to $c$, and the two relative fluctuations, $22/5c$ and $48/c\kappa L$, depend on the central charge in the same way.

For free bosons, a mode of frequency $\omega$ contains a number $N$ of quanta with the thermal distribution $p_N\propto e^{-2\pi\omega N/\kappa}$. Its mean value is $\langle N\rangle=n(\omega)$, Eq.~(\ref{eq:C-D}), and its variance is $\langle N^2\rangle-\langle N\rangle^2=n^2(\omega)+n(\omega)$. Equation~(\ref{eq:C-thermo}) is the sum of these variances over the modes, weighted with $\omega^2$. The two terms are known as the wave term and the particle term \cite{Pais82}. They are the integrals of the last term and of the cross term of Eq.~(\ref{eq:C-wick}), respectively, so that the Wick-ordered square accounts only for the wave term. For free fermions $N$ is $0$ or $1$, with mean value $n_F(\omega)=1/(e^{2\pi\omega/\kappa}+1)$ and variance $n_F(\omega)-n_F^2(\omega)$, and the Wick-ordered square reproduces only the quadratic term, which gives a negative variance (Appendix~\ref{app:W}). For $c$ free bosons and for $2c$ free Majorana fermions the three relative fluctuations are
\begin{equation}\label{eq:C-table}
\begin{array}{lccc}
 & \quad\text{thermodynamics, Eq.~(\ref{eq:C-thermo})}\quad & \quad\text{Eq.~(\ref{eq:C-local})}\quad & \quad W_{uu}\quad\\[3pt]
\text{bosons} & 48/c\kappa L & 22/5c & +2/c\\[2pt]
\text{Majorana fermions} & 48/c\kappa L & 22/5c & -1/2c
\end{array}
\end{equation}
One boson and two Majorana fermions have $c=1$, and the same correlation functions of the stress tensor in the thermal state, which are fixed by $c$. The thermodynamic result and Eq.~(\ref{eq:C-local}) coincide for the two theories, as they must for quantities defined in terms of the stress tensor alone. The Wick-ordered square gives $+2$ for the boson and $-1/2$ for the two fermions.

\textit{Point splitting.} Equation~(\ref{eq:C-local}) is the prescription (\ref{eq:A-normal}) in physical normalization, with the singular terms of Eq.~(\ref{eq:C-OPE}) evaluated at $u'$. In the thermal state it gives the same result for any $\alpha$. In a general state the prescriptions of Appendix~\ref{app:A} differ. In terms of the operators at the intermediate point $u_\lambda=u'+\lambda\Delta$, Eq.~(\ref{eq:C-OPE}) reads
\begin{equation}\label{eq:C-lambda}
T_{uu}(u)T_{uu}(u')=\frac{c}{8\pi^2\Delta^4}-\frac{T_{uu}(u_\lambda)}{\pi\Delta^2}-\frac{(1-2\lambda)\,\partial_uT_{uu}(u_\lambda)}{2\pi\Delta}-\frac{3-10\lambda(1-\lambda)}{20\pi}\,\partial_u^2T_{uu}(u_\lambda)+{:}\Lambda_{uu}{:}(u_\lambda)+O(\Delta),
\end{equation}
and if the singular terms of Eq.~(\ref{eq:C-lambda}) are the ones that are subtracted, the finite remainder is the operator ${:}T_{uu}^2{:}_\alpha={:}\Lambda_{uu}{:}-\frac{a}{20\pi}\partial_u^2{:}T_{uu}{:}$ with $\alpha=-\lambda(1-\lambda)$, Eq.~(\ref{eq:A-alpha}). Two prescriptions give variances that differ by a multiple of $\nabla_\pm\nabla_\pm\langle T_{\pm\pm}\rangle$, Eq.~(\ref{eq:variance}), which is fixed by the mean value and vanishes when $\langle T_{uu}\rangle$ is constant.


\begin{thebibliography}{99}
\bibitem{noteDFU} An alternative but equivalent expression was derived within a different approach in Ref.~\cite{DFU76}. See also Ref.~\cite{BirrellDavies}.
\bibitem{notePS} For a free massless scalar field a local expression for the renormalized fluctuations was obtained in Ref.~\cite{PerezSudarsky26}, with the square of the stress tensor defined by the Wick ordering of the four fields. Here we use instead the operator product expansion of the stress tensor, which applies to any conformal field theory and any central charge. The two definitions give different results (Appendix~\ref{app:W}).
\end{thebibliography}
\end{document}